\documentclass[aps,prl,twocolumn,showpacs,preprintnumbers,amsmath,amssymb]{revtex4}
\usepackage{epsfig}

\begin{document}
\title{Nonequilibrium phase transition due to communities isolation}
\author{Julian Sienkiewicz and Janusz A. Ho{\l}yst}
\affiliation{Faculty of Physics, Center of Excellence for Complex Systems Research, Warsaw University of Technology, Koszykowa 75, PL-00-662 Warsaw, Poland}
\date{\today}

\begin{abstract}
We introduce a simple model of a growing system with $m$ competing communities. The model corresponds to the phenomenon of defeats suffered by social groups living in isolation. A nonequilibrium phase transition is observed when at critical time $t_c$ the first isolated cluster occurs. In the one-dimensional system the volume of the new phase, i.e. the number of the isolated individuals, increases with time as $Z \sim t^3$. For a large number of possible communities the critical density of filled space equals to $\rho_c = (m/N)^{1/3}$ where $N$ is the system size. A similar transition is observed for Erd\H{o}s-R\'{e}nyi random graphs and Barab\'{a}si-Albert scale-free networks. Analytic results are in agreement with numerical simulations.
\end{abstract} 
 \maketitle

Recently physicists working on modeling of social phenomena are frequently touching the idea of {\it dissemination} and {\it competition} - especially in the case of language \cite{language}, culture \cite{axelrod} and opinions \cite{przegladowka}. The key subject can be posed as follows: how does the node internal variable change when it is influenced by others? The issue has occurred especially in Axelrod model of culture dissemination \cite{axelrod} or many sociophysics systems such as: Sznajd model \cite{sznajd},  voter model \cite{voter}, majority rule voting \cite{majority_rule}, social impact model \cite{social_impact} or bounded confidence models \cite{bounded_conf}. At this moment one should stress that the above mentioned models are bound to explore social effects of diffusion or adoption of node states. Such processes are usually running in accordance with the following scheme: one takes the state of its neighbor, provided that a set of rules is fulfilled. 

There is however also another, qualitatively different phenomenon -  {\it isolation} of surrounded social groups and resulted extinction of their members due to lack of communication with other groups of the same specie. In fact this issue  should be regarded as equally important as the previously mentioned dissemination or migration effects. The isolation and consequently the lack of communication among the groups belonging to the same community (and vice versa - no communication causing the isolation) might lead to severe disturbances in the society. One of them can be racial isolation (segregation) that can cause serious social problems \cite{seg_iso}. In fact the phenomenon of residential segregation has been studied in several physical papers \cite{seg} that in part follow the famous work of Schelling \cite{schelling} or other Ising-like approaches. In other situations the lack of social contact effects in increased mortality of seriously ill patients as compared to those that are not isolated \cite{health_iso}. Finally, recent research \cite{am_iso} shows that Americans suffer from social isolation due to dramatic decrease of number of discussion partners even with those that they share the closest relationship. Given the fact that such social phenomena as elections \cite{elections} or war \cite{war} are currently being examined using methods of statistical physics a quantitative model of social isolation could be a useful tool to predict blocking of voting districts or trapping of hostile troops during a wartime.

The key idea of our work can be presented in form of two questions: (i) what happens if instead of species spreading the interaction effects in species {\it isolation} and {\it extinction}? (ii) what are  consequences of the fact that sometimes a small group is capable to surround  and defeat a larger one? The first point comes as an effect of the observation that a group of people that is suddenly surrounded by people from opposite groups is  often defeated  by  enemies or opponents. It occurs because  the surrounded group is isolated and is not able to communicate with other group members who could support them, e.g. provide a military backup (it was a common case during many wars). To justify the second question one can think of Chinese game Go where, in some condition, one player can block the opponent using the amount of stones which is less than those that are just being surrounded. 

\begin{figure}[ht!]
\centerline{\epsfig{file=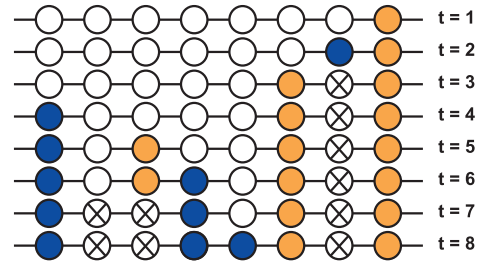,width=0.75\columnwidth}}
\caption{(Color online) An example of evolution in the chain consisting of 8 nodes. Open circles are empty sites, black and gray circles correspond to different communities. Isolated nodes are marked with a cross.} 
\label{fig:example}
\end{figure}

\begin{figure*}[ht!]
\centerline{\epsfig{file=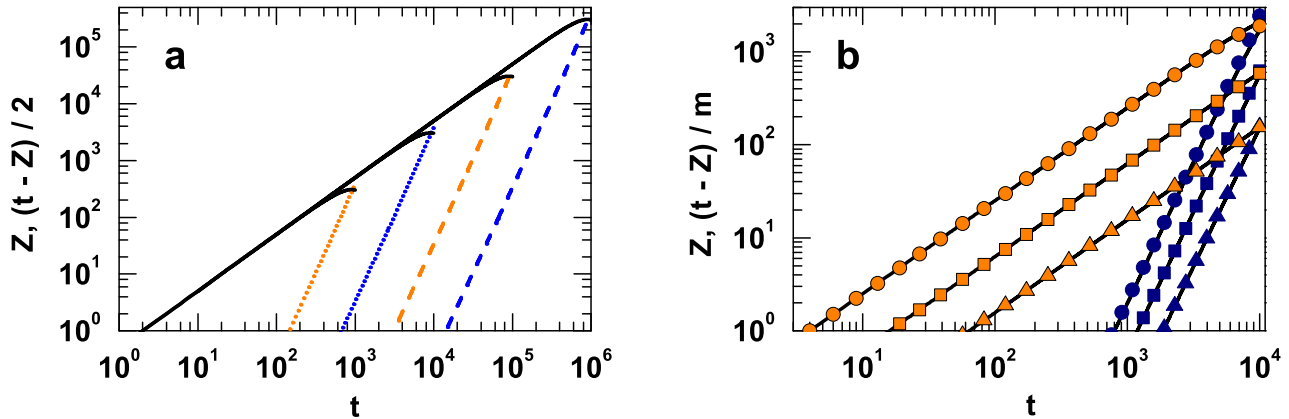,width=\textwidth}}
\caption{(Color online) {\bf (a)} Number of isolated nodes ($Z$, dotted and dashed lines) and not isolated nodes of each specie ($(t-Z)/2$, solid lines) versus time for different chain sizes $N$ (gray dotted - $10^3$, black dotted - $10^4$, gray dashed - $10^5$ and black dashed $10^6$). {\bf (b)} Number of isolated nodes ($Z$, filled symbols) and not isolated nodes of each specie ($(t-Z)/m$, open symbols) versus time for different number of species $m$ (circles - $m=4$, squares - $m=16$ and triangles - $m=64$). All simulations are for $N=10^4$ and the lines come form the solution of Eq. (\ref{eq:ukladm})}
\label{fig:evolution}
\end{figure*}

In this paper we impose these dynamical rules onto various regular and random networks. We start with a simple chain where sites are being filled with individuals belonging to two different species. Then we extend the model to a case of $m$ different species and finally we consider the Erd\H{o}s-R\'{e}nyi graphs and Barab\'{a}si-Albert scale-free networks.

Let us consider a chain of $N$ initially unoccupied nodes. In each time step one empty node is chosen randomly. Then, an internal variable ($\uparrow$) or ($\downarrow$) for this node is randomly selected. Both possibilities correspond to different species or communities and are drawn with the same probability. If a cluster of $n$ identical filled nodes (e.g. $\uparrow \uparrow \uparrow \uparrow$) is surrounded by individuals belonging to other community (e.g. $\downarrow \uparrow \uparrow \uparrow \uparrow \downarrow$), the nodes in the surrounded cluster are treated as extincted and can no longer interact with the rest of the chain, i.e. they will be not able to surround other clusters. The procedure is held until the chain is full, which happens in the $N$th time step. An example of complete evolution of the system is presented in Fig. \ref{fig:example}.

\begin{figure*}[ht!]
\centerline{\epsfig{file=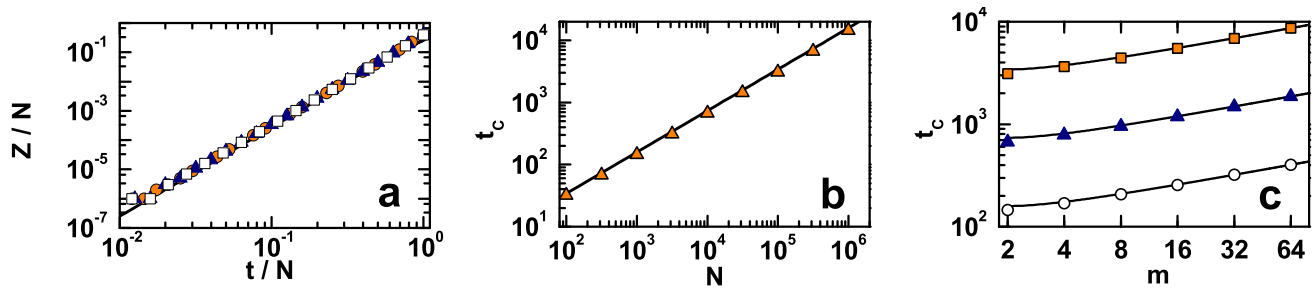,width=\textwidth}}
\caption{(Color online)  {\bf (a)} Data collapse for rescaled number of isolated nodes ($Z/N$) versus the rescaled time ($t/N$) observed for three different data sets - $N=10^4$ (circles), $N=10^5$ (triangles) and $N=10^6$ (squares). The curve (hardly visible) is obtained from Eq. (\ref{eq:zr}). {\bf (b)} The critical value of time $t_c$  for $m=2$ versus the chain size $N$. The line is Eq. (\ref{eq:tc}) while triangles are numerical simulations.  {\bf (c)} The critical value of time $t_c$ versus the number of species $m$ for different chain sizes $N=10^3$ (circles), $N=10^4$ (triangles) and $N=10^5$ (squares). Lines come from Eq. (\ref{eq:tcm}).}
\label{fig:tc}
\end{figure*}

Our main points of interest are: (i) the critical time $t_c$ when the first isolated cluster appears (ii) the number of isolated nodes for $t > t_c$.

Figure \ref{fig:evolution}a shows the number of isolated nodes $Z$ and the number of not isolated nodes of both species $(t-Z)/2$  as function of time for four different chain sizes: $10^3$, $10^4$, $10^5$ and $10^6$. In each case the number of isolated sites follows a power law $Z \sim t^{\alpha}$ with $\alpha$ exponent close to $3.0$ ($\alpha=3.09$ for $N=10^3$, $\alpha=3.06$ for $N=10^4$, $\alpha=3.05$ for $N=10^5$ and $\alpha=3.04$ for $N=10^6$).

The plots indicate that in this system we observe a nonequilibrium phase transition - after reaching a certain time of the evolution (after filling a specific number of nodes) a new phase emerges due to the occurrence of the first isolated cluster. The volume of this phase can be treated as the system order parameter. It grows up when one runs above the critical time $t_c$. Moreover  
it can be seen in Fig. \ref{fig:tc}a that data for different chain sizes collapses onto one curve after rescaling both $Z$ and $t$ axis by system size $N$. Figure \ref{fig:tc}b shows that the critical time $t_c$ of the first isolated node appearance grows with system size as $t_c \sim N^{\beta}$ where $\beta=0.664 \pm 0.001$.

In order to obtain the average number of isolated nodes, we have to sum all different possibilities of a cluster to become isolated. A single isolated site emerges either as an effect of a combination $\uparrow \downarrow \uparrow$ or $\downarrow \uparrow \downarrow$ in which the middle node is turned into isolated one. To express the total number of such nodes in the system ($Z_1$) we need to multiply the probability of the sum of those combinations by the number of such possibilities, that is $N-2$. Similarly the number of isolated sites coming from isolated clusters of size $n$ is:
\begin{equation}\label{eq:strzalki}
Z_n = n \left(N - n - 1 \right) \left[ {\Pr}^2(\uparrow) \prod_{i=1}^{i=n} \Pr(\downarrow) +  {\Pr}^2(\downarrow) \prod_{i=1}^{i=n} \Pr(\uparrow) \right],
\end{equation}
where $n=1,2,\ldots$. As the examined system is symmetric (i.e. $\Pr(\uparrow) = \Pr(\downarrow)$), taking into account that at time $t$ there are already $Z$ isolated nodes the average probability of finding a certain specie at time $t$ is $(t-Z)/(2N)$. Since $Z = \sum_{i=1}^{i=n}Z_i$ we obtain after short algebra:
\begin{equation}\label{eq:sum}
Z = 2 \sum_{n=3}^{n= \infty} \left( n-2 \right) \left( N-n+1 \right)\left(\frac{t-Z}{2 N} \right)^{n}.
\end{equation}
or $Z = (t - Z)^3 / (2N - t + Z)^2$. Solving this equation leads to  
\begin{equation}\label{eq:zr}
Z_r = \frac{1}{6}\left[ 5t_r - 4 + (8 + 16 t_r - t_r^2){u_r}^{-1}-u_r \right]
\end{equation}
with $u_r = (\sqrt{3(16-24t_r + 39t_r^2 -2t_r^3)}-80 + 84 t_r -24 t_r^2)^{1/3}$ where $t_r=t/N$ and $Z_r=Z/N$. The formula is universal for any value of chain size - all data should collapse on this curve, as it can be seen in Fig. \ref{fig:tc}a. 
If $Z \ll t \ll N$, what is acceptable for the most part of the evolution, then 
Eq. (\ref{eq:sum}) leads to  $Z \approx t^3 / (4N^2)$ i.e. the number of isolated nodes should increase as $t^3$. This fact is in agreement with the numerical experiment. This approximated formula can be also used to calculate the critical time $t_c$ at which the first isolated node appears in the chain. Putting $Z=1$ we get a simple expression for the critical time
\begin{equation}\label{eq:tc}
t_c = \left(2 N \right)^{2/3}.
\end{equation}
This result is consistent with the value of the $\beta$ parameter observed in the numerical data.

We can easily extend the previously described model of two competing species onto a case where the number of species is $m \geq 2$. Similarly to the two-species case, in each time step a type of specie is drawn from the uniform distribution $\langle 1, m \rangle$ and placed in a random, unoccupied place in the chain. The isolated nodes are formed from a cluster of identical species surrounded by other {\it identical} species.

An example of the evolution of the extended model is presented in Fig. \ref{fig:evolution}b. Like in the case of $m=2$, the number of isolated nodes follows a power-law $Z \sim t^{\gamma}$ with $\gamma$ exponent close to $3$ ($\gamma=3.05$ for $m=4$, $\gamma=3.00$ for $m=16$ and $\gamma=3.01$ for $m=64$).

The analytical approach in the case $m > 2$ is identical to the case $m=2$ except for two things. First, there are $m$ different species which can be isolated. Each of those $m$ species can be isolated in $m-1$ ways, therefore instead of factor $2$ we should put $m(m-1)$ in front of each equation in the set of equations (\ref{eq:strzalki}). Second, larger number of species results in the change of probability of finding a specific specie - in the extended model it is equal to $(t-Z)/(mN)$. Thus Eq. (\ref{eq:strzalki}) has now the form
\begin{equation}\label{eq:ukladm}
Z_n =  n \cdot m (m - 1) (N - n - 1)\left(\frac{t - Z}{mN}\right)^{n+2}
\end{equation}
where $n=1,2,\ldots$ . Following an identical algebra as in the case of two-species model, we arrive at a self-consistent equation for the number of isolated nodes $Z = (m - 1)(t - Z )^{3} / (mN - t + Z)^{2}$ which is exactly algebraically solvable. The solution fits to the numerical data quite well (see Fig. \ref{fig:evolution}b) and as before one can approximate it with $Z \approx (m-1)t^3/(mN)^2$. This proves that the increase of isolated nodes follows the same rule as in the two-specie case, i.e. $t^3$. The critical time $t_c$ at which the first isolated node appears is
\begin{equation}\label{eq:tcm}
t_c = \left( \frac{m^2}{m-1} \right)^{1/3} N^{2/3}.
\end{equation}
which, once again, is consistent with the numerical data (see Fig. \ref{fig:tc}c).

The form of Eq. (\ref{eq:tcm}) gives us the opportunity to spot the interplay between the only two parameters of the model - the length of the chain $N$ and the number of species $m$. If $m \gg 1$ the Eq. (\ref{eq:tcm}) can be rewritten in a form of $t_c=(mN^2)^{1/3}$, what leads to the critical density  of filled nodes $\rho_c=(m / N)^{1/3}$. The obvious conclusion from this relation is that when the chain becomes larger the critical density gets smaller and in the thermodynamical limit vanishes completely. Then, if we would like to maintain a constant value of $\rho_c$ we should require $m/N = \rm const$. In other words, it is possible to prevent the convergence of critical density to zero by making the number of species proportional to the chain's length. 

\begin{figure}[ht!]
\centerline{\epsfig{file=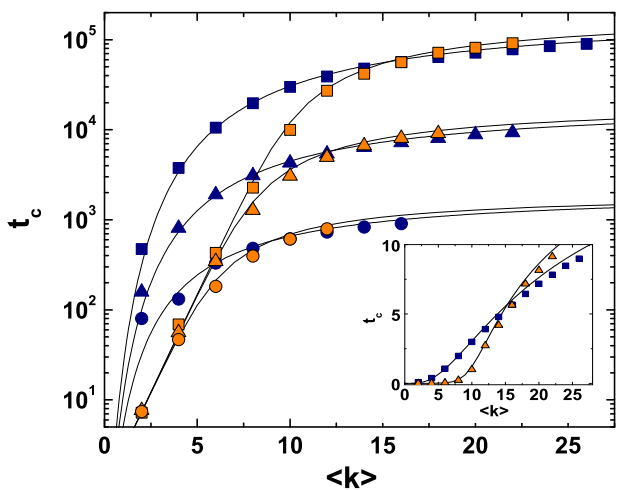,width=\columnwidth}}
\caption{(Color online) Logarithmic plot of the critical time $t_c$ for Erd\H{o}s-R\'{e}nyi graphs (empty symbols) and Barab\'{a}si-Albert networks (full symbols) for different networks sizes: $N=10^3$ (circles), $N=10^4$ (triangles) and $N=10^5$ (squares). Symbols are numerical simulations and lines come from Eqs. (\ref{eq:tcer}) and (\ref{eq:tcba}). The inset shows $t_c$ in linear scale for $N=10^5$ in case of ER graph (triangles) and BA network (squares) - $t_c$ is rescaled by a factor of 10000.}
\label{fig:ba}
\end{figure}

The critical time $t_c$ for $m=2$ can be also found in the case of two- and three-dimensional cubic lattices and, what is more important, for any random network characterized with a specific degree probability distribution $p(k)$ ($k$ is the number of links of a given node). The general formula for number of single isolated nodes is
\begin{equation}\label{eq:tcpk}
Z_1 = 2N\sum_{k=0}^{k=\infty}p(k)x^{k+1},
\end{equation}
where $x=t/(2N)$. In order to obtain the critical time, we require that $Z_1=1$ and solve this equation for $t$. In case of regular lattices we have $t_c=(2N)^{z/(z+1)}$, where $z$ is number of neighbors. We have calculated the critical time for two most popular types of complex networks: Erd\H{o}s-R\'{e}nyi random graphs \cite{er} and Barab\'{a}si-Albert scale-free networks \cite{ba}. In case of ER graphs characterized by degree distribution $p(k)=e^{-\langle k \rangle} {\langle k \rangle}^k / k!$ the critical time $t_{ER}$ can be expressed as
\begin{equation}\label{eq:tcer}
t_{ER} = \frac{2N}{\langle k \rangle} W \left( \frac{ e^{\langle k \rangle} \langle k \rangle}{2N} \right).
\end{equation}
where $W(x)$ is Lambert W-function. For BA network (degree distribution $p(k)=\frac{1}{2}{\langle k \rangle}^2 k^{-3}$) we get
\begin{equation}\label{eq:tcba}
t_{BA} \approx \left( \frac{\langle k \rangle}{4} \right)^{\frac{2}{\langle k \rangle + 2}} \left( 2N \right)^{\frac{\langle k \rangle}{\langle k \rangle + 2}}.
\end{equation}
The obvious condition for avoiding an isolated node in the system is $t_c > N$. It leads to the following inequalities for different networks: $N > 2^z$ for regular lattices, $N > 2^{\frac{\langle k \rangle-4}{2}}\langle k \rangle$ for BA networks and $N > e^{\langle k \rangle/2}$ for ER graphs. The above described results are shown in Fig. \ref{fig:ba}.

{\it Conclusions} - In this work we proposed a simple approach to model communities isolation in growing societies. The numerical simulations, fully supported by analytical approach show that  a critical time $t_c$ a nonequilibrium phase transition takes place and a new phase consisting of surrounded clusters emerges. In the case of one-dimensional system the number of isolated nodes rises with time as  a power-law with exponent $\gamma=3$. The scaling is universal i.e. it   depends neither on the chain's length $N$ nor on the number $m$ of possible species.  An analytic form for the critical time $t_c$ is found and for large $m$  this time scales as $t_c=(m N^2)^{1/3}$. The phenomenon has been also observed for higher dimensional systems as well as for Erd\H{o}s-R\'{e}nyi random graphs and Barab\'{a}si-Albert scale-free networks.

JS and JAH acknowledge a support from  the EU Grant {\it Measuring and Modelling Complex Networks Across Domains} - MMCOMNET (Grant No. FP6-2003-NEST-Path-012999) and from Polish Ministry of Education Science (Grant No. 13/6.PR UE/2005/7). JS is thankful to Axel Leijonhufvud for useful comments.

\end{document}